\documentclass[conference]{IEEEtran}
\IEEEoverridecommandlockouts

\usepackage{cite}
\usepackage{amsmath,amssymb,amsfonts}
\usepackage{algorithmic}
\usepackage{textcomp}
\usepackage{xcolor}
\usepackage{caption}
\usepackage{subcaption}
\usepackage{hyperref}

\usepackage{flushend}
\usepackage{fancyhdr}
\usepackage{algorithmic}
\usepackage{graphicx}
\usepackage{textcomp}
\usepackage{xcolor}
\usepackage{tabularx}
\usepackage{colortbl}
\usepackage{tikz}
\usepackage{enumitem}
\usepackage{booktabs}
\usetikzlibrary{arrows, chains, positioning, shapes.geometric, shapes.symbols}
\usetikzlibrary{fadings,calc,patterns}
\usepackage{caption}
\usepackage{subcaption}
\usepackage{booktabs}   
\usepackage{subcaption} 

\usepackage{graphicx}

\begin{document}

\title{GraphBinMatch: Graph-based Similarity Learning for Cross-Language Binary and Source Code Matching}

\author{\IEEEauthorblockN{Ali TehraniJamsaz, Hanze Chen, Ali Jannesari}
\IEEEauthorblockA{\textit{Iowa State University, Ames, Iowa, USA}}
\IEEEauthorblockA{\{tehrani, hanzech, jannesar\}@iastate.edu}
\\[-3.0ex]
}

\maketitle

\begin{abstract}
Matching binary to source code and vice versa has various applications in different fields, such as computer security, software engineering, and reverse engineering. Even though there exist methods that try to match source code with binary code to accelerate the reverse engineering process, most of them are designed to focus on one programming language. However, in real life, programs are developed using different programming languages depending on their requirements. Thus, cross-language binary-to-source code matching has recently gained more attention. Nonetheless, the existing approaches still struggle to have precise predictions due to the inherent difficulties when the problem of matching binary code and source code needs to be addressed across programming languages.

In this paper, we address the problem of cross-language binary source code matching. We propose GraphBinMatch, an approach based on a graph neural network that learns the similarity between binary and source codes.
We evaluate GraphBinMatch on several tasks, such as cross-language binary-to-source code matching and cross-language source-to-source matching. We also evaluate our approach's performance on single-language binary-to-source code matching.
Experimental results show that GraphBinMatch outperforms state-of-the-art significantly with improvements as high as 15\% over the F1 score. 
\end{abstract}

\maketitle


\maketitle

\section{Introduction}
\label{Introduction}
Binary code is a collection of instructions that can be executed by computing systems directly, whereas source code, which programmers write, is readable and understandable.
Binary-to-source code matching is a technique to evaluate the likeliness of binary code and source code.
This is an important aspect of many security software engineering tasks, such as vulnerability \cite{yarlagadda2020approach} and malware detection\cite{yang2017bmxnet} and reverses engineering\cite{miyani2017binpro,shahkar2016matching}.

Typically, when it comes to matching a binary code to a source code, we either want to find the match for the binary file or the source code file. For example, when we have a binary code fragment, it would be helpful to retrieve its similar source code snipped, which can be used in a reverse engineering task. The retired source code snipped enables researchers to understand what a binary code fragment does.
From the other aspect, if we have a source code snipped with a vulnerability, matching it to a binary code form helps to identify whether the vulnerability exists in the binary file.

Existing approaches try to measure the semantic similarity between binary and source code. However, most works focus on matching binary to binary or source code to source code \cite{zhao2018deepsim, yu2020order, yuan2019b2sfinder}.
Binary-to-source code is a non-trivial task as two modalities are involved: binary and source codes.
Lately, recent works have been trying to measure the similarity between binary and source code; however, they fall short in matching binary-to-source code across different programming languages. Recently, Gui \textit{et al.}\cite{gui2022cross} proposed a transformer-based neural network to learn the similarity of binary and source code across programming languages; they use Intermediate Representation (IR) as input data for their model. However, they treat IR as a sequence of tokens; thus, their model struggles to learn the similarity across programming languages.

In this paper, we present GraphBinMatch. An approach based on a graph neural network to learn the semantic similarities between binary and source code. Unlike previous approaches, Binary and source code are tread as graphs by leveraging three types of flows: Control flow, data flow, and call flow. Presenting binary and source codes as graphs helps the neural network model better learn the code's structure and semantics since the source code contains specific structures, unlike natural language text. 

Experimental results show that GraphBinMatch outperforms state-of-the-art approaches by increasing F1 from 0.65 to 0.79, recall from 0.59 to 0.82, and precision from 0.73 to 0.76.

Overall, the major contributions of this paper are:
\begin{enumerate}
    \item Formulating the problem as learning the similarities between graphs.
    \item Developing a special graph neural network as the backbone of GraphBinMatch to learn the similarity of graphs.
    \item Evaluation of GraphBinMatch on a comprehensive set of tasks.
    \item Effectiveness of the approach not just for cross-language but also single-language.
    \item Up to 15\% improvement in comparison to state-of-the-art approach.
\end{enumerate}

The rest of the paper is structured as follows:
We first formulate the binary-source matching problem in section \ref{Formulating the problem}. Then, in section \ref{Approach}, our proposed approach is outlined and explained. 
Experimental setups are discussed and explained in section \ref{experimental_setup}.
Section \ref{evalution} presents the evaluation results, followed by section \ref{Discussions} in which we discuss some of the insights.
Next, in section \ref{Related Works}, we provide an overview of related works, and lastly, section \ref{Conclusion and Future Works} concludes the paper and explains the future works.

\section{Formulating the Problem}
\label{Formulating the problem}
We formulate cross-language binary code-matching detection as follows:
Given two programs $P_a$ as binary and $P_b$ as source written in two different programming languages, we aim to train a deep learning model to learn the function $\gamma$, which can predict whether the two input programs are a binary-code matching pair or non-binary-code matching pair.\par

The training set consists of triples $(P_a, P_b, y_{ab})$ where $y_{ab}$ is the label. We consider all pairs of binary and source programs collected from the same coding task in the dataset as positive samples and label them as 1, indicating that they are binary-source matches. Conversely, we consider all pairs of binary and source programs generated from different tasks in the dataset as negative samples and label them as 0, indicating that they are non-binary-source matches.

\begin{equation}
    \gamma(P_a,P_b)= 
\begin{cases}
\label{train_formula}
    1,& \text{if $P_a$ and $P_b$ are matching}\\
    0,              & \text{otherwise}
\end{cases}
\end{equation}
\section{Approach}
\label{Approach}
\begin{figure*}[htbp]
\centering
\includegraphics[width=\textwidth]{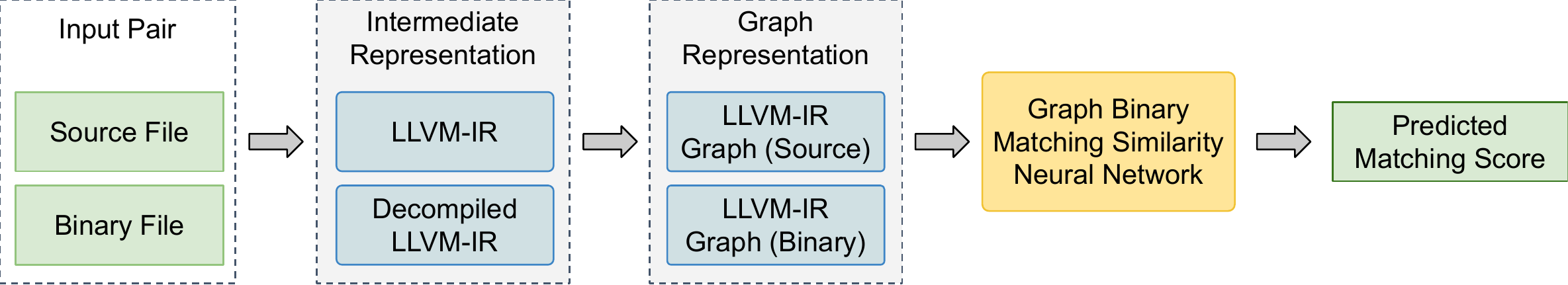}
        \caption{
        The overview of our proposed approach.
        }
        \label{overall_workflow}
\end{figure*}

This section will discuss our proposed approach for identifying binary-source matching pairs across programming languages. The overall workflow is shown in Figure \ref{overall_workflow}. The input to GraphBinMatch consists of two files, a source file and a binary file written in different programming languages. The first step is to compile the source file and decompile the binary file using the respective language-specific front-ends and tools, to produce Intermediate Representations (IR) that are language-independent. Next, we create graphs capturing the programs' control, data, and call flow. We treat these graphs as heterogeneous graphs to better model the different types of nodes and edges. An edge in the graphs represents the relationships between two nodes. These heterogeneous graphs are inputs to our Graph Binary Matching Similarity Neural Network (GraphBinMatch).

\subsection{Intermediate Representation}
As mentioned, the first step in our proposed approach is to convert the files from the source language and binary to an intermediate representation. Intermediate representation (IR) is an intermediate form that modern compilers use to represent  programs. IR provides a more straightforward abstraction. It allows multiple stages of transformation and analysis in the compiler to generate the target code more efficiently. Lowering programs to intermediate representation is a common compiler technique to simplify and optimize code. In GraphBinMatch, we first convert input files to \texttt{LLVM IR}. Using LLVM IR, our approach can represent input files in a language-independent format, allowing for easier comparison of code written in different programming languages. \par

We use two front-ends, \texttt{JLang} \footnote{\href{https://polyglot-compiler.github.io/JLang}{https://polyglot-compiler.github.io/JLang}} and \texttt{Clang} \footnote{\href{https://clang.llvm.org}{https://clang.llvm.org}}, to convert Java and C++ programs to LLVM IR, respectively. \texttt{JLang} supports Java up to version 7, while Clang-5.0 converts C++ programs to LLVM IR. It is worth mentioning that different versions of LLVM can produce slightly different IRs, so using the same version of Clang as the one being used inherently by \texttt{JLang} helps to have more similarities between Java and C++ programs.

Our proposed approach uses RetDec\footnote{ \href{https://github.com/avast/retdec}{https://github.com/avast/retdec}} to generate LLVM IR from binary executables.
RetDec is an open-source machine-code decompiler that can generate LLVM IR from binary executables. It supports various architectures, including x86, ARM, MIPS, and PowerPC. RetDec uses instruction parsing, data-flow analysis, and control-flow reconstruction techniques to reverse-engineer the binary code into LLVM IR. 

\subsection{Graph Generation}
While there exist various approaches \cite{ben2018neural, venkatakeerthy2020ir2vec, sui2020flow2vec} to represent \texttt{LLVM IR} for deep learning models, recently, it has been shown that presenting programs as graphs can help deep learning models to learn the characteristics of programs more effectively \cite{allamanis2022graph, allamanis2017learning, cummins2020programl}.
Following the recent success in presenting programs as graphs, we also create graphs using LLVM IR files. In particular, we use \texttt{ProGraML} \cite{cummins2020programl} to generate a graph for each of the LLVM IR files in our dataset. \texttt{ProGraML} extracts information from LLVM IR and constructs a graph consisting of different types of nodes (i.e., instruction, variable and constant) and edges (i.e., control flow, data flow, and call flow). These graphs capture programs' structural and semantic information and can be used as inputs to machine learning models for various program analysis tasks, including code similarity detection. These graphs must be encoded and passed to a graph neural network designed to learn the similarities among matching pairs and dissimilarities among non-matching pairs.

\subsection{Node Feature Embedding}
To generate node embeddings, we split the process into two parts: preprocessing of the source code and processing within the model itself. During source file preprocessing, we create a feature vector for each node using the \texttt{full\_text} or \texttt{text} of the node's attributes in the graph generated by \texttt{ProGraML}. The \texttt{full\_text} attribute represents the complete LLVM-IR instruction for the node, whereas the \texttt{text} attribute only contains the corresponding instruction type. For instance, in the instruction \texttt{\%16 = load i32\, i32\* \%15\, align 8}, means the integer pointer \%15 is loaded and stored in temporary variable \%16. We can only obtain the instruction type from the \texttt{text} attribute. However, using the \texttt{full\_text} attribute, we can determine that the load instruction is responsible for loading an integer pointer. This additional information can be used to better train the embedding layer and improve the model's accuracy.\par

We discovered that not all nodes have the \texttt{full\_text} attribute during the preprocessing process, as some nodes only represent compiler configurations that only have \texttt{text} section. To address this issue, we use the \texttt{text} attribute as a fallback option when the \texttt{full\_text} attribute is unavailable.

The ProGraML paper itself uses \texttt{text} attribute of the nodes as the feature of the nodes; however, later in the experimental result section, we will see that using the \texttt{full\_text} attribute of the nodes improves the prediction of the GraphBinMatch.
Moreover, we use the \texttt{position} property provided by \texttt{ProGraML} as the edge feature. The \texttt{position} property in \texttt{ProGraML} contains the edge position information. For instance, for a valid LLVM-IR instruction such as \texttt{\%result = add i32 \%1, \%2}, \texttt{ProGraML} generates an edge corresponding to \texttt{\%1} and an edge corresponding to \texttt{\%2}. In this example, the position of the first edge is 0, and the position of the second edge is 1.

Finally, we use the tokenizer to tokenize those LLVM-IR instructions to create the final node features, which we then add to the graph.
Tokenizer is also able to map each token to a corresponding integer number. Therefore we would have a sequence of integer numbers representing an LLVM-IR instruction for a node. This sequence of integer numbers is considered as the feature of the node.
In the conversion process, we convert all LLVM-IR variables, such as \texttt{\$12}, to a special token named \texttt{[VAR]}. By utilizing the tokenizer and carefully selecting the truncation and padding lengths, we ensure that our feature vectors are informative and can be effectively used in subsequent modeling steps.

LLVM-IR instructions have varying lengths; therefore, the feature set of each node in our graph will have its own length. To solve this problem and make sure that all nodes have the same length for their features, We use the average length of all nodes' feature vectors rounded up to the nearest power of 2 as the final intercept length. For example, if the average length of all LLVM-IR instructions in the dataset is 50 after tokenization, the final truncation length is 64. All feature vectors with a length of 64 will be truncated, and all feature vectors with a length less than 64 will be padded with \texttt{[PAD]} token.\par

\begin{figure*}[h]
\centering
\includegraphics[width=\textwidth]{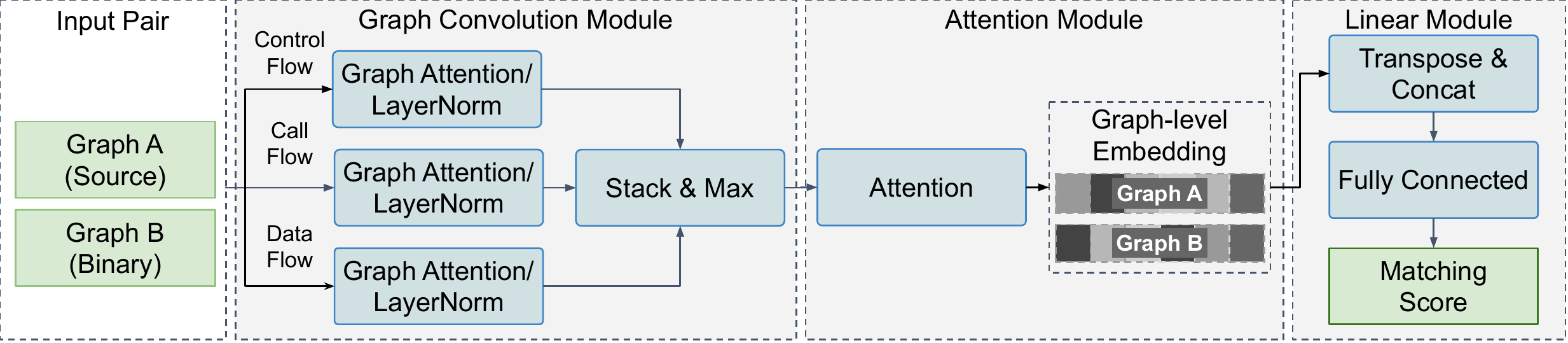}
        \caption{
        Structure of Graph Binary Matching Similarity Neural Network (GraphBinMatch).
        }
        \label{model_architecture}
\end{figure*}

\subsection{Graph Binary Matching Similarity Neural Network}
Once the preprocessing is finished, we will pass the graphs to our Graph Binary Matching Similarity Neural Network (GraphBinMatch) for further processing. GraphBinMatch is built upon SimGNN \cite{bai2019simgnn}, but we have made several modifications to tailor it for heterogeneous graph similarity tasks, which will be discussed in the next sections. \par

The architecture of GraphBinMatch can be seen in Figure \ref{model_architecture}. To train GraphBinMatch, we create data points consisting of three items: \textbf{Source File A}, \textbf{Binary File B}, and \textbf{Label}. 
We use an embedding layer as the first layer of GraphBinMatch. This layer processes the feature vector of each node and tries to embed the nodes by learning the embedding space so that similar nodes would have similar embedding. The embedding layer increases the dimensionality of the entire feature vector to two dimensions. We utilize the max operation to reduce the two-dimensional feature vector to a single dimension. This one-dimensional feature vector serves as input to the graph convolution layer.\par
In the following subsections, we will provide a detailed explanation of each component of GraphBinMatch.

\subsubsection{\textbf{Heterogeneous Convolution}}~\par
GraphBinMatch is designed to take in the graph representation of two files and predict whether they match. The Convolution layer is constructed to support the heterogeneous graph using the \texttt{HeteroConv} wrapper from the \texttt{pytorch-geometric} library. As previously discussed, our Heterogeneous Graph representation has three types of relationships. This layer includes three separated \texttt{GATv2Conv}\cite{brody2022how} layers to model each one of the relationships. The outputs of the \texttt{GATv2Conv} layers are stacked together, and then the element-wise maximum values are computed to have a latent representation of the nodes. After each \texttt{GATv2Conv}, we include additional \texttt{LayerNorm} to stabilize training and prevent overfitting.


\subsubsection{\textbf{Attention}}~\par
In GraphBinMatch, we incorporate an attention layer similar to the one introduced in \cite{bai2019simgnn}. The attention layer operates by taking node embeddings from the previous layer and passing them through an attention mechanism. We use a global graph embedding vector $c \in \mathbb{R}^D$ where $D$ is the same dimension as the nodes' embedding dimension. $c$ is created by averaging node embeddings passed through a non-linear transformation. This global embedding vector $c$ contains the overall structure and information of the graph.\par

To calculate the attention for a given node $n_i$, we compute the inner product of $c$ and the latent representation of $n_i$. The intuition behind this approach is that nodes that are more similar to the overall context of the graph should receive higher attention weights. Finally, the graph-level embedding is created by computing the weighted sum of all nodes. Here, the nodes' weights refer to the nodes' attention values. 

\subsubsection{\textbf{Fully Connected Layer}}~\par
Once the graph-level embeddings on the two files are computed, the two vectors are concatenated and transposed to be fed to the fully connected layer to have the final prediction. In our study, we use two fully connected layers. After the first fully connected layer, a normalization layer is also applied to stabilize the learning process. Additionally, we introduce an extra dropout layer before the last linear layer to increase nonlinearity and regularization in the model.

In the next section, we evaluate GraphBinMatch and compare its results with the state-of-the-art approach.
\section{Experimental Setup}
In this section, we evaluate GraphBinMatch and present the results and address four research questions.
\label{experimental_setup}
\subsection{Tasks and Research Questions}
We evaluate GraphBinMatch on cross-language and single-language tasks. For each one of these tasks, we use a specific dataset.
We define the research questions as follows
\begin{itemize}
    \item\textbf{RQ1:} Can graph-based representation and GNNs outperform transformer-based models in cross-language binary source code matching problems? 
    \item\textbf{RQ2:} Does GraphBinMatch provide consistent results when applied in a single-language context with different optimization levels?
    \item\textbf{RQ3:} How is the performance of GraphBinMatch across different compilers?
    \item\textbf{RQ4:} Does GraphBinMatch perform well in terms of source-to-source matching?
\end{itemize}
\subsection{Dataset Statistic}
\textbf{Cross Language Code Matching}. To evaluate GraphBinMatch for the cross-language binary-source matching task, we use the CLCDSA dataset \cite{nafi2019clcdsa}, which consists of source code files collected from two programming competition websites: AtCoder and Google CodeJam. These programming competition websites feature multiple tasks, and the dataset contains source code as solutions to the tasks in various programming languages. We have selected c, c++, and java as the languages for our study to assess the effectiveness of our model in learning binary-source code similarities across different programming languages. \par

To ensure that our dataset has a balanced distribution of positive and negative samples, we consider valid solutions to the same task as matching pairs and valid solutions to different tasks as non-matching pairs. We discard any file that is not compilable.
Following the baseline paper, we adopt the same train, validation, and test split ratio, which is a ratio of 6:2:2 to split the dataset.

As mentioned, in this study, we utilize JLang\footnote{\href{https://github.com/polyglot-compiler/JLang}{https://github.com/polyglot-compiler/JLang}} and clang-5.0 as compilers for all java source code files. JLang is an open-source compiler front-end that compiles java source code into the corresponding LLVM-IR.
This compiler is used to generate all corresponding LLVM-IR expressions for Java files. 
To convert LLVM-IR to binary executable files, \texttt{Clang-5.0} is used.
Throughout the experiments, 0z is set as the default optimization level of the compiler unless the optimization level is explicitly mentioned. 
To convert all binary executables to their LLVM-IR equivalent, we use an open-source RetDec decompiler.\par

\begin{table*}[htbp]
    \centering
    \caption{Dataset Statistics}
    \label{dataset-stats}
    \begin{tabular}{cccccc}
            \hline&
            \multicolumn{1}{l}{Languages} &
            \multicolumn{1}{l}{\# Sources} &
            \multicolumn{1}{l}{\# LLVM-IR} &
            \multicolumn{1}{l}{\# Binary Files} &
            \multicolumn{1}{l}{\# Decompiled LLVM—IR} \\ \hline
            \multicolumn{1}{c}{} & C    & 15605 & 13929 & 14370 & 13929 \\
            CLCDSA               & C++  & 16676 & 15375 & 15766 & 15589 \\
                                 & Java & 19836 & 15124 & 17072 & 15124 \\ \hline
            POJ-104              & C++  & 52000 & 38598 & 38598 & 37909 \\ \hline
    \end{tabular}%
\end{table*}

\textbf{Same Language Code Matching} For the task of detecting matching within the same programming language, we use the POJ-104 dataset \cite{mou2016convolutional} \footnote{\href{ https://drive.google.com/uc?id=0B2i-vWnOu7MxVlJwQXN6eVNONUU}{https://drive.google.com/uc?id=0B2i-vWnOu7MxVlJwQXN6eVNONUU}}. This dataset comprises C++ solutions submitted by 500 students for 104 different programming problems from an online judge system (OJ) that serves an educational purpose. We compile the C++ source code files into LLVM-IR and binary executable using different optimization levels of \texttt{clang} and \texttt{gcc}. To decompile the binary executable to its corresponding LLVM-IR representation, we utilized RetDec similarly for all the binary executables.

Table\ref{dataset-stats} shows the statistics of the two datasets.

\subsection{Baselines}
As discussed, We evaluate the performance of GraphBinMatch for two different matching tasks: binary-source matching and source-source matching.
To assess the effectiveness of GraphBinMatch, we compare it against BinPro\cite{miyani2017binpro}, B2SFinder\cite{yuan2019b2sfinder}, and XLIR for binary-source matching. 

BinPro is a tool that aims to tackle the challenge of identifying similarities between source and binary code even when the compiler or optimization level used is unknown. To this end, BinPro employs machine learning techniques to compute the best code properties for determining binary-to-source code similarity. These code properties are then extracted and computed using static analysis tools to match binary and source codes with a bipartite matching algorithm.

B2SFinder detects binary code clones by inferring seven traceable features in binary and source code. It employs a weighted feature-matching algorithm capable of handling different features and calculating the weights of code feature instances based on their specificity and frequency of occurrence. 

XLIR is a transformer-based neural network model which is currently state-of-the-art for binary-source code matching. XLIR, as its name suggest, also uses LLVM IR. To embed the tokens in LLVM IR, XLIR leverages a pre-trained BERT model. The model first pre-trains the neural network using a large external LLVM-IR corpus with masked language modeling (MLM)\cite{sinha2021masked} as the pre-processing step. This step is aimed at learning meaningful representations of the LLVM-IR tokens. Once the tokens are embedded, XLIR maps them into a common space, and the LLVM-IR representations are learned jointly using a ternary loss function. This approach allows XLIR to match binary source code across different programming languages.

For source-source matching, in addition to XLIR, we also compare the results against LICAA\cite{vislavski2018licca}. \textbf{LICCA} works on the source code and extracts semantic and syntactic characteristics of programs to identify matching pairs.

\subsection{Experiment Setup}
GraphBinMatch is built using \texttt{Pytorch-Geometric}\footnote{\href{https://pytorch-geometric.readthedocs.io}{https://pytorch-geometric.readthedocs.io}}, a powerful library for developing deep learning models for graphs.
For optimizing the learning parameters of GraphBinMatch, Adam \cite{kingma2014adam} Optimizer is used with a learning rate of $6.6e^{-5}$, and Binary Cross Entropy is used as the loss function.
We use  GATv2\cite{brody2022how} as the graph convolution layer as discussed earlier.
The hyper-parameters of GraphBinMatch are tuned using RayTune\footnote{\href{https://docs.ray.io/en/latest/tune/index.html}{https://docs.ray.io/en/latest/tune/index.html}}.
The HuggingFace tokenizer\footnote{\href{https://huggingface.co/docs/transformers/index}{https://huggingface.co/docs/transformers/index}} is used to tokenize the LLVM-IR instructions which serve as nodes' features in our graph representation. 

GraphBinMatch comprises a PyTorch Embedding layer with a dimension of 128 to embed the tokens, five GATv2 graph convolution layers with a dimension of 256, and a HuggingFace GPT tokenizer with a max number of 2048 vocabularies. In the GraphBinMatch model, we use \texttt{LeakyReLU} as our activation function, except the last linear layer is followed by a \texttt{Sigmoid} function.
We train GraphBinMatch using four a100 NVIDIA GPUs with 80GB VRAM and Intel Xeon 6140 CPU with 128GB RAM.

\subsection{Evaluation Metrics}
\begin{table}[htbp]
    \begin{center}
        \begin{tabular}{ccc}
        \hline
        Parameter                   & Prediction         & Actual \\ \hline
        True Positive(\textbf{TP})  & Matching              & Matching           \\
        True Negative(\textbf{TN})  & Non-matching          & Non-matching       \\
        False Positive(\textbf{FP}) & Matching              & Non-matching       \\
        False Negative(\textbf{FN}) & Non-matching          & Matching           \\ \hline
        \end{tabular}
    \end{center}
    \caption{}
    \label{avaluate_matrix}
\end{table}

In the realm of code matching detection, precision (P), recall (R), and F1-scores (F1) are commonly used to evaluate the performance and accuracy of the models. These three criteria are derived from four measures: true positive (TP), true negative (TN), false positive (FP), and false negative (FN). The definitions of these measures can be found in Table\ref{avaluate_matrix}.

\textbf{Precision} is often used to describe the accuracy of a model's positive prediction, that is, the accuracy for pairs identified as matching. It is defined as the proportion of true matching pairs out of all the matching pairs predicted by the model, as shown in Equation \ref{precision-equation}.\par
\begin{equation}
    \label{precision-equation}
    P = \frac{TP}{TP+FP}
\end{equation}

\textbf{Recall} is defined as the percentage of matching pairs in the dataset that the model correctly predicts as matching pairs, which is the case described by Equation \ref{recall-equation}.
\par
\begin{equation}
    \label{recall-equation}
    R = \frac{TP}{TP+FN}
\end{equation}
\textbf{F1-Score} is commonly used to evaluate the performance of models. F1-Score is defined as the harmonic mean of the precision and recall values, which is the case described by Equation \ref{recall-equation}
\begin{equation}
    \label{f1-equation}
    F1 = \frac{2PR}{P + R}
\end{equation}
\section{Experimental Results}
\label{evalution}
In this section, we address the research questions.
\subsubsection{\textbf{RQ1: Can graph-based representation and GNNs outperform transformer-based models in cross-language binary source code matching problem?}}~\par
\begin{table*}[htbp]
    \centering
    \caption{Performance of cross-language binary-mating task (Threshold at 0.5).}
    \label{tab:cross_binary_matching_result}
    \begin{tabular}{l|ccc|ccc}
        \hline& 
        \multicolumn{3}{c|}{C/C++ binary code with Java source code} & \multicolumn{3}{c}{Java binary code with C/C++ source code} \\ \hline
                          & Precision & Recall & F1   & Precision & Recall        & F1   \\ \hline
        BinPro            & -         & -      & -    & 0.36      & 0.37          & 0.36 \\
        B2SFinder         & -         & -      & -    & 0.35      & 0.41          & 0.38 \\
        XLIR(LSTM)        & 0.62      & 0.53   & 0.57 & 0.55      & 0.51          & 0.53 \\
        XLIR(Transformer) & 0.73      & 0.59   & 0.65 & 0.68      & 0.55          & 0.61 \\
        GraphBinMatch         & 0.75      & 0.73   & 0.74 & 0.75      & \textbf{0.78} & 0.77 \\
        GraphBinMatch(Tokenizer) & \textbf{0.76}      & \textbf{0.82}      & \textbf{0.79}      & \textbf{0.76}         & 0.77         & \textbf{0.77}        \\ \hline
    \end{tabular}    
    \textsuperscript{*Because of the limitations of BinPro and B2Sfinder in handling Java source code, we cannot provide test results for Java as source code} 
    \textsuperscript{*In this table and the following ones, the results of the other tools are quoted from the baseline paper.}
\end{table*}

To answer the question of RQ1, we trained our model using the CLCDSA dataset.
We evaluated the performance of GraphBinMatch in the binary-source matching task by comparing it against the aforementioned baselines, as shown in Table\ref{tab:cross_binary_matching_result}. We used LLVM-IR from binary C/C++ programs and java source code as GraphBinMatch's input in our experiments. The results show the effectiveness of GraphBinMatch with the precision, recall, and F1 scores achieving 0.76, 0.82, and 0.79, respectively. This represents over 20\% improvement compared to the baseline paper. To further strengthen the robustness of our results, we conducted an additional experiment using LLVM-IR from Java binary and C/C++ source code. This yielded satisfactory results, with precision, recall, and F1 scores of 0.76, 0.77, and 0.77, respectively, which exceeded the baseline by 25\%.\par

Table \ref{tab:cross_binary_matching_result} reveals a performance gap between using binary Java source code and binary C/C++ code for the same task and dataset. This gap may be attributed to certain differences between the LLVM-IR obtained by decompiling and the one obtained from source code, as GraphBinMatch struggles to comprehend. The decompiled LLVM-IR is not always identical to the source code, and we attribute this difference to two primary factors. Firstly, the decompiled code may not always have the exact data type or the correct shape of arrays for array types. Secondly, the decompilation process often involves speculation and assumptions, resulting in variations in the control flow generated by decompiling binaries. Combining these factors within the binary file causes differences in the LLVM-IR.

\subsubsection{\textbf{RQ2: Does GraphBinMatch provide consistent results when applied in a single-language context with different optimization levels?}}~\par
\begin{table}[htbp]
    \centering
    \caption{Performace of single language binary matching task (Threshold at 0.5).}
    \label{tab:same_lang_vs_baseline}
    \begin{tabular}{l|ccc}
        \hline
        \textbf{}         & Precision & Recall & F1   \\ \hline
        BinPro            & 0.38      & 0.42   & 0.40 \\
        B2SFinder         & 0.43      & 0.46   & 0.44 \\
        XLIR(LSTM)        & 0.67      & 0.72   & 0.44 \\
        XLIR(Transformer) & 0.85      & 0.86   & 0.85 \\
        GraphBinMatch     & 0.88      & 0.86   & 0.87 \\ \hline
    \end{tabular}
\end{table}

\begin{table}[htbp]
    \centering
    \caption{Same language binary matching result from different optimization level}
    \label{tab:same_lang_at_different_opt_level}
    \begin{tabular}{c|ccc|ccc}
        \hline
        \textbf{} & \multicolumn{3}{c|}{clang-10.0} & \multicolumn{3}{c}{gcc-9.4} \\ \hline
        \textbf{} & Precision   & Recall   & F1     & Precision  & Recall  & F1   \\ \hline
        O0        & 0.88        & 0.86     & 0.87   & 0.87       & 0.86    & 0.87 \\
        O1        & 0.87        & 0.88     & 0.88   & 0.89       & 0.85    & 0.85 \\
        O2        & 0.86        & 0.82     & 0.84   & 0.87       & 0.83    & 0.85 \\
        O3        & 0.86        & 0.83     & 0.85   & 0.84       & 0.81    & 0.83 \\
        Oz        & 0.90        & 0.85     & 0.87   & 0.87       & 0.87    & 0.87 \\ \hline
    \end{tabular}
\end{table}

We can confidently answer yes to this research question. For this study, we used a different dataset than CLCDSA due to insufficient data for the same language. As shown in Table \ref{dataset-stats}, CLCDSA has a maximum of only 15589 C++ source codes available, whereas the POJ-105 dataset provides 37909 source codes. This replacement of the dataset provides  1.5 times more code. POJ-105 is also the dataset that has been used in the baseline paper as well. Table \ref{tab:same_lang_vs_baseline} reveals that GraphBinMatch outperforms the baseline paper regarding precision, recall, and F1 scores, with scores of 0.88, 0.86, and 0.87, respectively, indicating that GraphBinMatch is more proficient at detecting binary-source matching of the same language.
This table shows that the transformer-based model (XLIR) can perform quite well as the LLVM-IR from the same language since these IRs show more similarities. Despite the good XLIR(Transformer) scores on a single language, GraphBinMatch still outperforms XLIR(Transformer).
\par

To verify the robustness of GraphBinMatch in different optimization scenarios, we evaluated its performance using clang on the same dataset but with four different optimization levels (O0, O1, O2, O3, Oz). Table\ref{tab:same_lang_at_different_opt_level} shows that GraphBinMatch achieves consistent performance across different optimization levels. This indicates that the model's performance is not dependent on a specific compiler or optimization level and can provide reliable results under varying conditions.\par

By examining the performance of GraphBinMatch with different optimization levels of the same compiler, we can obtain some insights regarding RQ2. We suspect that higher optimization levels would result in more aggressive optimizations by the compiler, such as control flow tuning. These optimizations would provide the decompiler with additional assumptions and speculations, resulting in an increased difference between LLVM-IR from both the source code and the binary executable. As shown in Table \ref{tab:same_lang_at_different_opt_level}, we observe that the precision, recall, and F1 scores used to measure model performance gradually decrease as the optimization level increases. This suggests that more aggressive optimizations can slightly affect the model's performance in code matching detection.\par

\subsubsection{\textbf{RQ3: How is the performance of GraphBinMatch across different compilers?}}~\par
GraphBinMatch showed consistent performance across different optimization levels of the same compiler; in this subsection, we want to investigate further to see whether it could also provide similar performance across different compilers. This is an important consideration for real-world applications where the compiler used to generate a binary executable may be unknown. To test this, we used \texttt{clang} to generate LLVM-IR from the source code, but \texttt{gcc} to generate the binary executable of the POJ-104s dataset compared to RQ2. We kept the same settings as in RQ2 to convert the binary executable to the corresponding LLVM-IR using RetDec. Table \ref{tab:same_lang_at_different_opt_level} presents the performance of GraphBinMatch when generating binary executable using \texttt{gcc}. The results show that GraphBinMatch performs relatively the same when using different compilers.

We suspect the better performance figures of \texttt{gcc} than clang because the used decompilers tend to provide more information when compiling the C++ code generated by \texttt{gcc}. This is supported by our analysis, which shows that the average size of LLVM-IR generated by binary executables compiled with \texttt{clang} is about $10,769.9$ bytes, while the average size of those compiled with \texttt{gcc} is about $18,525.2$ bytes. This means at the decompilation stage, the size of \texttt{gcc} compiled binary files is approximately 70\% larger than those compiled with a \texttt{clang}. Such a significant difference in size will likely affect GraphBinMatch's ability to accurately detect binary matching.

\subsubsection{\textbf{RQ4: Does GraphBinMatch perform well in terms of source-to-source matching?}}~\par
\begin{table*}[htbp]
    \centering
    \caption{Cross language source matching result}
    \label{tab:corss_lang_src_marching_result}
    \begin{tabular}{c|ccc|ccc|ccc}
        \hline
        \textbf{} & \multicolumn{3}{c|}{GraphBinMatch} & \multicolumn{3}{c|}{XLIR (LSTM)} & \multicolumn{3}{c}{XLIR (Transformer)} \\ \hline
        \textbf{}     & Precision     & Recall        & F1            & Precision & Recall & F1   & Precision     & Recall & F1   \\ \cline{2-10} 
        C vs Java     & \textbf{0.77} & \textbf{0.80} & \textbf{0.78} & 0.62      & 0.51   & 0.56 & 0.75          & 0.55   & 0.63 \\
        C++ vs Java   & 0.76          & \textbf{0.82} & \textbf{0.79} & 0.65      & 0.53   & 0.58 & \textbf{0.77} & 0.57   & 0.66 \\
        C/C++ vs Java & 0.81          & 0.73          & 0.78          & -         & -      & -    & -             & -      & -    \\ \hline
    \end{tabular}
\end{table*}

In the source-source matching task, we evaluate the performance of GraphBinMatch using the CLCDSA dataset and compare it with the baseline paper. The experimental setup was the same as in RQ1, except that we used the LLVM-IR generated by \texttt{JLang} as input to GraphBinMatch. We test three language combinations: C/C++ vs. Java, C vs. Java, and C++ vs. Java. Table \ref{tab:corss_lang_src_marching_result} shows that GraphBinMatch outperformed the baseline paper by about 20\% regarding precision, recall, and F1 scores. GraphBinMatch can also effectively detect source code matching across different programming languages.

\subsection{Varying Threshold}
\begin{figure}[ht]
\includegraphics[width=0.5\textwidth]{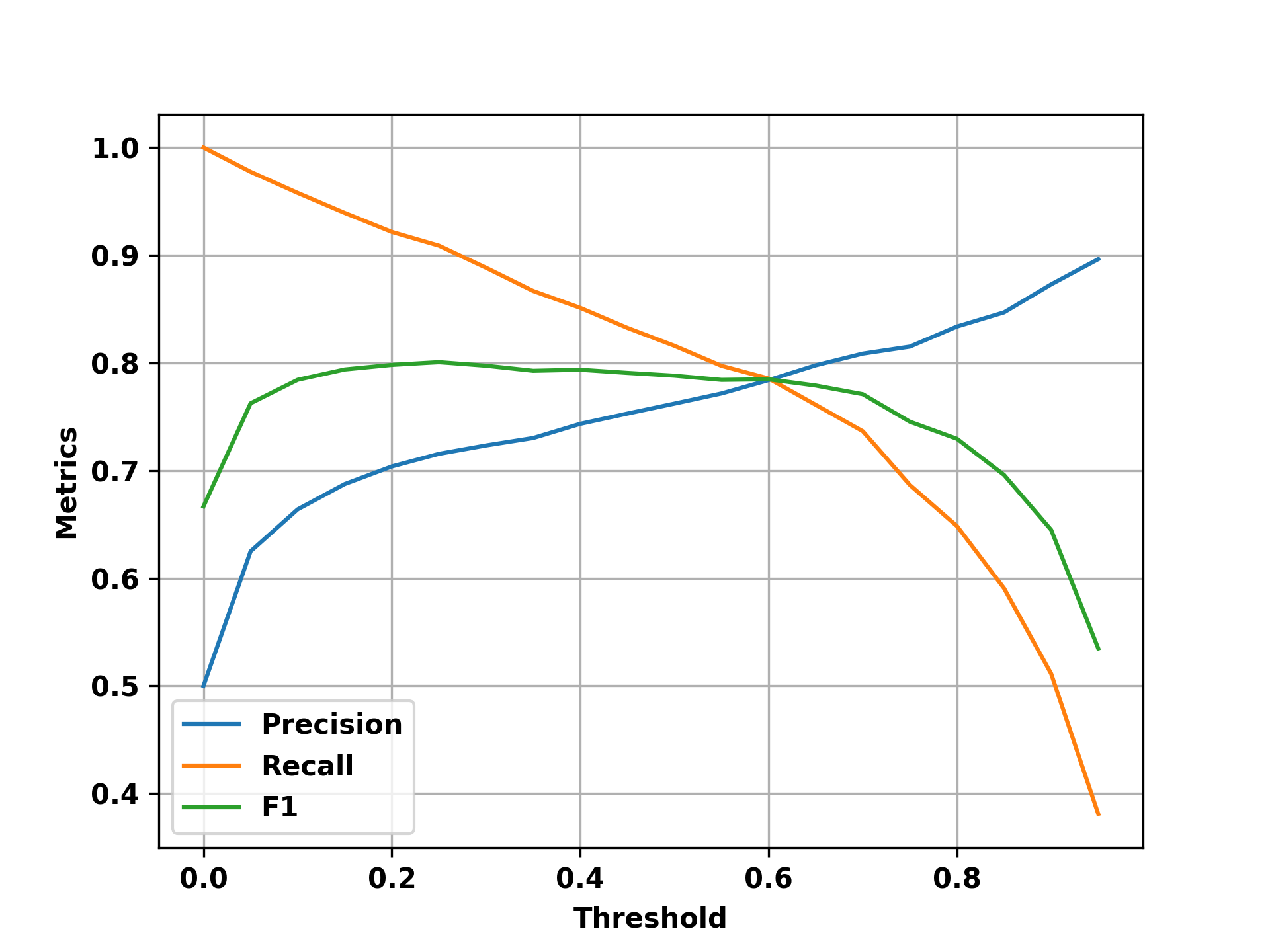}
\caption{ [Higher is better] Varying the threshold results in different scores for precision, recall, and f1.}
\label{varying-threshold}
\end{figure}

In the previous subsection, we mentioned that the threshold for GraphBinMatch was set to 0.5. We experimented with varying thresholds to investigate the effect of different thresholds on precision, recall, and F1 scores. Figure \ref{varying-threshold} shows the different scores that can be achieved by varying the threshold. Based on our findings, a threshold value of 0.2 would provide a slightly better F1 score. However, we also observed that using this threshold would significantly decrease accuracy to as high as 7\%, making it impractical to use for optimal F1 score to find the best threshold. As a result, we chose to use a threshold value of 0.5, which we considered to be a more reasonable default threshold for our study. A smaller threshold yields higher recall since GraphBinMatch will predict all pairs as matching pairs. On the other hand, a larger threshold will result in higher precision, as GraphBinMatch predicts all pairs as non-matching pairs. Depending on the use case, a user of GraphBinMatch can manually set the threshold or let GraphBinMatch decide the best threshold based on the given metric (i.e., precision, recall, F1).

\section{Discussions}
In this section, we discuss some challenges that can affect the performance of GraphBinMatch.
\label{Discussions}
\begin{table}[htbp]
    \centering
    \caption{Statistics for the number of nodes in test set}
    \begin{tabular}{ccc}
        \hline
        Type           & Mean & Median \\ \hline
        True Positive  & 1506 & 864    \\
        False Positive & 2133 & 1303   \\
        True Negative  & 2573 & 1680   \\
        False Negative & 2293 & 1289   \\ \hline
    \end{tabular}
    \label{fig:stat-number-nodes}
\end{table}

\subsection{Investigating why GraphBinMatch may fail}
Our experiments revealed that our model occasionally misidentifies pairs of matching code fragments. After reviewing the mispredicted samples individually, we found that most false positives occur because of a large gap in the sizes of the LLVM-IR fragments. We calculated and analyzed the number of nodes in the test set graphs and obtained Table \ref{fig:stat-number-nodes}. The table shows that the difference in the mean and the median number of nodes predicted by our model is much larger for the false positive set than for the true positive set. The median difference in the number of nodes between the two sets is nearly 50\%.\par 

After conducting an in-depth analysis, we found two main reasons for these false positives. The first reason is that the gap between LLVM-IR is larger than the tolerance of GraphBinMatch. For example, some code snippets may use sorting methods provided by the standard library, while others may implement their own sorting methods internally. GraphBinMatch may not effectively recognize that the method calls of the standard library are equivalent to the sorting methods the authors themselves have implemented in their code. Additionally, the template mechanism in C++ can impact GraphBinMatch since many of the C++ standard libraries are published as templates. This means that templates are also compiled as a part of LLVM-IR, which causes struggles for GraphBinMatch to recognize that the compiled template code is equivalent to standard library calls in Java.

\begin{figure*}[htbp]
    \begin{subfigure}{.48\textwidth}
    \centering
    \includegraphics[width=0.9\textwidth]{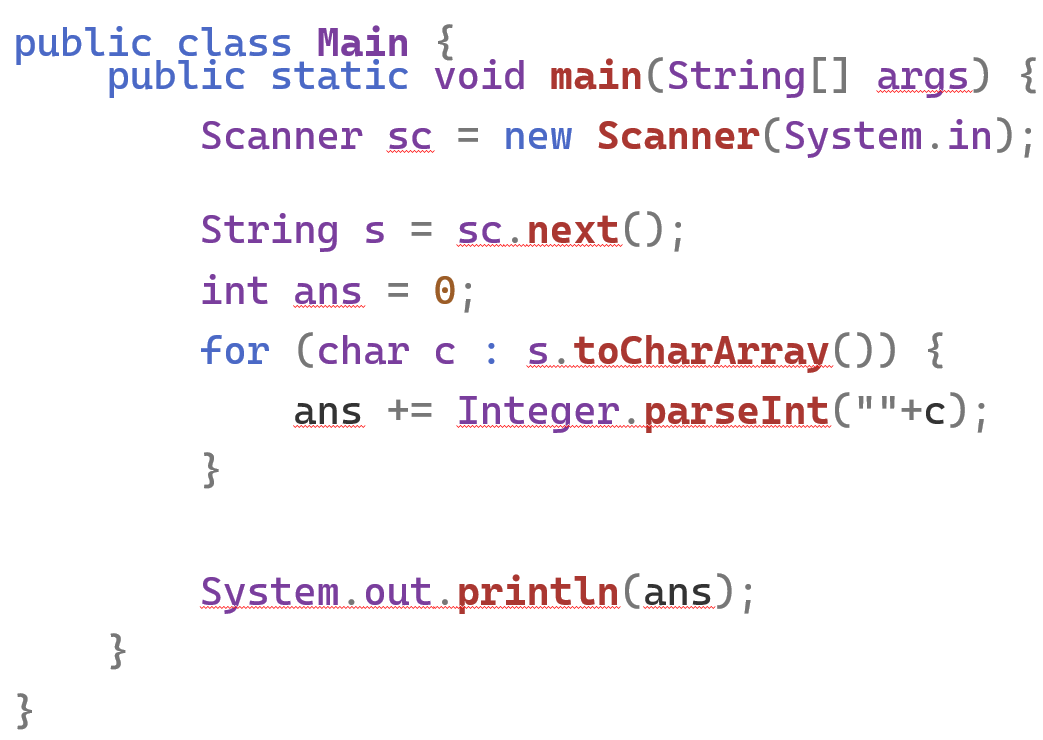}
    \caption{Samples from Java}
    \label{sf1}
  \end{subfigure}
  \begin{subfigure}{.48\textwidth}
    \centering
        \includegraphics[width=0.9\textwidth]{figures/java.png}

    \caption{Samples from c++}
    \label{sf2}
  \end{subfigure}
    \caption{An example of false negative case}
    \label{fig:false-negative}
\end{figure*}

The second scenario involves language differences between Java and C++ and different usage habits among their respective user communities, which can result in significant discrepancies in the LLVM-IR even when both are converted to compile as LLVM-IR. These discrepancies are rooted in the varying habits of different programming language users. For example, Figure \ref{fig:false-negative} displays a matching pair that, despite their similarity, generate IR graphs with vastly different sizes: the Java-generated IR graph has 330 nodes and 660 edges, whereas the C++-generated IR graph has only 65 nodes and 115 edges. Such discrepancies can cause challenges for GraphBinMatch to recognize matching pairs, as the model may not account for the impact of language usage habits on the IR generation process, leading to incorrect results.

\subsection{Extending GraphBinMatch to other programming languages}
This study introduced a novel approach to detecting cross-language binary-code matching using graph neural networks (GNNs). A crucial component of our approach is using LLVM IR to measure the similarities between two programs written in different languages. By leveraging LLVM IR, we can capture high-level information about the code independent of the underlying programming languages.

While our approach is designed to be cross-language, it relies on the availability of compiler front-ends. One must use the corresponding compiler front-end to extend our approach to support additional programming languages to produce LLVM IR. Once the IR is generated, our approach remains the same. Our method is flexible enough to support any programming language compiled into LLVM IR.

\subsection{How tokenizer and different token embedding influence the result}
\begin{table}[htbp]
    \caption{Performance of different LLVM-IR embedding techniques for same-language binary matching and cross-language binary matching}
    \label{tab:ablation_result}
    \begin{tabular}{l|ccc|ccc}
        \hline
                                                & \multicolumn{3}{|c|}{Cpp vs Cpp } & \multicolumn{3}{|c}{Cpp/C vs Java } \\ \hline
                                                & Precision        & Recall        & F1         & Precision         & Recall        & F1          \\ \hline
        text                                    & 0.86             & 0.83          & 0.85       & 0.75              & 0.73          & 0.74        \\
        full\_text & 0.89             & 0.87          & 0.88       & 0.84              & 0.75          & 0.79        \\\hline
    \end{tabular}
\end{table}

In this section, we conduct an experimental study using the CLCDSA dataset with the same settings and preprocessing methods as in RQ1. Our objective is to investigate the impact of various token embedding techniques and tokenizers on the performance of GraphBinMatch. Specifically, we aim to explore the effects of different embedding methods and tokenization schemes on the ability of GraphBinMatch to match binary and source code pairs across multiple programming languages or the same languages. By examining these factors, we hope to better understand how to optimize the performance of GraphBinMatch and improve its ability to detect code clones in diverse language pairs. \par

First, we aim to evaluate the impact of using the \texttt{full\_text} property versus \texttt{full\_text} for the source-binary matching task. In the approach section of our paper, we mentioned that using the \texttt{full\_text} property provided by \texttt{ProGraML} could lead to better performance than using \texttt{full\_text} alone. As shown in Table \ref{tab:ablation_result}, we observe that using the \texttt{full\_text} attribute improves the performance for both cross-language binary-source matching tasks and same-language binary-source matching tasks. Notably, for cross-language binary-source code tasks, using the \texttt{full\_text} attribute provides more significant improvement than binary-source code matching tasks written in the same language. We believe this result is because by exposing more information to the model, we can better bridge the understanding gap of the model on cross-language tasks.

\section{Related Works}
\label{Related Works}
Code similarity detection has recently gained considerable attention with the advent of research tools and machine learning models. These advancements provide new opportunities for research in this field. In code similarity detection, the similarity of source code is generally classified into four levels, as outlined in \cite{allamanis2017learning}:\par
\begin{itemize}
    \item \textbf{Type \uppercase\expandafter{\romannumeral1}}: This is also called \textit{Exact Clone}. The source code can be identical, with only the indentation, comments, and code layout modified.
    \item \textbf{Type \uppercase\expandafter{\romannumeral2}}: This is also called \textit{Parameterized clone}. The source code's structure and syntactic are similar except for some variable names, method names, and data types modified in addition to those mentioned in Type \uppercase\expandafter{\romannumeral1}.
    \item \textbf{Type \uppercase\expandafter{\romannumeral3}}: Compared to Type \uppercase\expandafter{\romannumeral2}, Type \uppercase\expandafter{\romannumeral3} code clone involves modification of statements, but the functional similarity is maintained.
    \item \textbf{Type \uppercase\expandafter{\romannumeral4}}: The structure between two code fragments is syntactically and structurally different, but both code fragments perform similar behavior for the same input.\par
\end{itemize}\par

Research on code similarity detection is broadly divided into two research directions: algorithm-based and machine learning-based. Each of these two different directions will be described in detail below:
\subsection{Algorithm based Approaches}
The algorithm-based code clone detection is based on lexical or semantic analysis, so most do not have a good detection result for Type \uppercase\expandafter{\romannumeral4} code clones, or even when Type \uppercase\expandafter{\romannumeral3} gaps are too large or too frequent. Most of the algorithms are only applicable to Type \uppercase\expandafter{\romannumeral1}, Type \uppercase\expandafter{\romannumeral2}, and part of Type \uppercase\expandafter{\romannumeral3} code clones. Also, their results for cross-language code clone detection are often unsatisfactory. \par

Research projects like CCFinderSW\cite{8305997} and SourcerCC\cite{10.1145/2884781.2884877} use a token-based approach to detect code clones. Each of these tools uses its own implementation of a lexical analyzer to convert the source code into a token stream which serves as a basis for analysis and similarity score measurement using different algorithms. However, because of the lack of lexical information abstraction, it is difficult for these algorithms to obtain and understand the inherent semantics.\par

Semantic analysis-based code clone detection tools try to solve this problem. DECKARD\cite{4222572} proposes a tree-based code clone detection scheme: it innovatively uses feature vectors to describe the structure of the syntax tree and clusters the clones by the Euclidean distance of the feature vectors.\par

Traditional algorithm-based code clone detection methods are often difficult to transfer to cross-language clone detection, resulting in very few studies addressing this problem. Even if the relevant studies propose corresponding solutions, they are limited by various restrictions and cannot address real-world requirements. Two of the most representative studies are LICCA\cite{8330250} and CLCMiner\cite{cheng2017clcminer}. LICCA detects code clones by attempting to convert different languages into uniform expressions, which can only cover cases where two pieces of code have a similar structure and syntactic elements. CLCMiner uses an NLP approach to view the source code but relies on the modification records of the source code.\par

\subsection{Machine Learning Approaches}
With the advancements in machine learning and graph neural networks, new ideas have emerged to solve the code clone detection problem. With the introduction of various code cloning datasets, it has become possible to use machine learning to solve such problems. Like CLCDSA\cite{nafi2019clcdsa} and BigCloneBench\cite{svajlenko2021bigclonebench}, they both contain code clones with Type \uppercase\expandafter{\romannumeral4} similarity. Many recent studies have tried to replace traditional algorithms with machine learning models to achieve better accuracy and gain the ability to compare across different languages. \par

CLDH\cite{ijcai2017p423} first proposed a hash method to detect whether two pieces of code in the same language are clones.
Their model accepts raw code fragments as input, encodes the abstract syntax tree using specific rules, and transforms it into a vector representation of the source code using an LSTM network. However, their encoding approach focuses more on the structure of the AST without considering other semantic information like node type. \par

In addition to tree-based methods, methods like CCLearner\cite{li2017cclearner} and C4\cite{tao2022c4} use token-based code clone detection: One of the first studies to use tokens as input to a code clone detection model is CCLearner. It classifies tokens into eight categories, calculates the similarity score for each category separately, and then uses the similarity of vectors to calculate the similarity between code fragments. This study concluded that more similarities in feature vectors imply a higher probability that a code pair is a code clone. This study only captures token and partial syntax level information but not the structure of the code. Their paper stated that CCLearner does not have good detection performance for Type \uppercase\expandafter{\romannumeral4}/Type \uppercase\expandafter{\romannumeral3} clones due to this reason.\par

Another token-based research is C4\cite{tao2022c4}. Their study used a pre-trained Bert model called codebert\cite{feng2020codebert} to embed the model's inputs. Since CodeBERT supports encoding multiple languages, C4 also supports cross-language code clone detection. Their study uses raw source code as input and first use CodeBERT to encode the raw source code to obtain a feature vector, and then their model learns and classifies the feature vector to obtain the final similarity score. They also implemented Contrastive Learning to increase the usage of the dataset. However, C4 only supports encoding the first 512 tokens because of the limitation of CodeBERT, affecting the final applicability of C4. \par

The current academic focus is mainly on source code clone detection, but Gui \textit{et al.}'s XLIR has shifted the attention back to the problem of cross-language binary source code matching. It is common for software applications to be written in different programming languages to meet various requirements and computing platforms. Therefore, detecting binary source code clones across multiple programming languages can be beneficial in practical scenarios. For instance, when vulnerable binary code is detected, it becomes necessary to retrieve relevant source code fragments of all possible programming languages written for better vulnerability assessment. We use XLIR as the baseline paper; more details on its methodology and performance are provided in the \ref{evalution} section.
\section{Conclusion and Future Works}
This paper introduced GraphBinMatch, a novel cross-language binary-source matching model that uses LLVM IRs and Heterogeneous graphs to learn the similarities. The model takes LLVM IR as input and can handle three types of code-matching detection tasks: cross-language binary matching detection, single-language binary matching detection, and cross-language source matching detection. It should be noted that GraphBinMatch is not limited to these specific languages. It can support any programming language with a compiler front end to generate LLVM IR.

Experimental results show that GraphBinMatch is effective and outperforms all three state-of-the-art tools. Specifically, our approach outperforms these tools regarding precision, recall, and F1 score, which indicates that it excels in identifying cross-language code matches.

In future work, we aim to extend GraphBinMatch to more programming languages, incorporating additional compiler front-ends to generate LLVM IR for these languages.
Moreover, since our approach is data-driven, we are collecting more source code files to increase the size of the dataset, which will believe will improve the prediction of GraphBinMatch.
By incorporating these improvements, we believe that GraphBinMatch will become a more powerful tool for code-matching detection and benefit software developers in maintaining and improving their code base.
\label{Conclusion and Future Works}

\bibliographystyle{IEEEtran}
\bibliography{text/bibliography}

\end{document}